\documentclass{article}

    \PassOptionsToPackage{numbers, compress}{natbib}

    \usepackage[final]{neurips_2025}

\usepackage[utf8]{inputenc} %
\usepackage[T1]{fontenc}    %
\usepackage{hyperref}       %
\usepackage{url}            %
\usepackage{booktabs}       %
\usepackage{amsfonts}       %
\usepackage{nicefrac}       %
\usepackage{microtype}      %
\usepackage{xcolor}         %

\usepackage{wrapfig}
\usepackage{graphicx}
\usepackage{tcolorbox}
\usepackage{amsmath}
\usepackage{tabularx}

\newtcolorbox{insights}[1][]{%
    colback=blue!5!white, %
    colframe=blue!75!black, %
    sharp corners, %
    toprule=0pt, %
    bottomrule=0pt, %
    leftrule=1.2pt, %
    rightrule=1.2pt, %
    #1 %
}

\title{Will LLMs Scaling Hit the Wall? Breaking Barriers via Distributed Resources on Massive Edge Devices}

\author{Tao Shen$^{1,}$\thanks{Equal contribution}, Didi Zhu$^{1,}$\footnotemark[1], Ziyu Zhao$^{1,}$\footnotemark[1], Zexi Li$^{1,}$\thanks{Corresponding author}, Chao Wu$^{2,}$\footnotemark[2], Fei Wu$^{1,}$\footnotemark[2] \\
$^{1}$College of Computer Science and Technology, Zhejiang University, China \\
$^{2}$School of Public Affairs, Zhejiang University, China \\
\texttt{\{tao.shen, didi.zhu, ziyuzhao.cs, zexi.li, chao.wu, wufei\}@zju.edu.cn}}

\begin{document}

\maketitle

\begin{abstract}
The remarkable success of foundation models has been driven by scaling laws, demonstrating that model performance improves predictably with increased \textit{training data} and \textit{model size}. 
However, this scaling trajectory faces two critical challenges: the exhaustion of high-quality public data, and the prohibitive computational power required for larger models, which have been monopolized by tech giants. 
These two bottlenecks pose significant obstacles to the further development of AI.
In this position paper, we argue that leveraging massive distributed edge devices can break through these barriers. 
We reveal the vast untapped potential of data and computational resources on massive edge devices, and review recent technical advancements in distributed/federated learning that make this new paradigm viable.
Our analysis suggests that by collaborating on edge devices, everyone can participate in training large language models with small edge devices.
This paradigm shift towards distributed training on edge has the potential to democratize AI development and foster a more inclusive AI community. 
The project page is available at \url{https://tao-shen.github.io/Distributed-LLM-Edges/}
\end{abstract}

\section{Introduction}

\textbf{Scaling laws} \citep{kaplan2020scaling,hoffmann2022training} have been fundamental to the remarkable success of foundation models, demonstrating a predictable relationship between performance and the expansion of model parameters and training data. These laws have guided the development of increasingly powerful models, from BERT \citep{devlin2018bert} to GPT-4 \citep{openai2023gpt4}, showing that performance improvements can be achieved through systematic scaling of both model size and training data \citep{brown2020language,chowdhery2022palm}. However, the continued application of these scaling laws requires ever-increasing amounts of data and computational resources, pushing the boundaries of what is currently feasible \citep{patterson2021carbon}.

\textbf{Public data} has been the primary \textit{fuel} driving AI development forward. 
This field has witnessed an exponential growth in data requirements, from the early success of MNIST \citep{lecun1998mnist} with its 70,000 handwritten digits to ImageNet's revolutionary impact with 14 million labeled images \citep{deng2009imagenet}. 
This trajectory has continued with modern large language models (LLMs) like GPT \citep{openai2023gpt4}, LLaMA \citep{touvron2023llama3}, and DeepSeek \citep{liu2024deepseek} series, which are trained on trillions of tokens.
Recent evidence that LLaMA 3.1's smallest model (8B) \citep{meta2024llama3.1} trained on 15 trillion tokens, outperforms LLaMA 2's largest model (70B) \citep{touvron2023llama} trained on 2 trillion tokens (despite being $10 \times$ smaller in model size, the $7 \times$ increase in training data leads to superior performance), demonstrates the paramount importance of data scaling \citep{raffel2020exploring,deeplearningai2024federated}. 
However, we are witnessing a concerning trend of data exhaustion, where high-quality public data sources are becoming exhausted \citep{lee2021dedup}. \cite{villalobos2024position} argues that human-generated public text data cannot sustain scaling beyond this decade. 
While recent efforts advocate for training larger models with synthetic data \citep{chen2024diversity}, AI-generated content may fail to yield performance improvements \citep{wenger2024ai}, also risks polluting public data sources \citep{fang2024bias}. 
Moreover, stricter data privacy regulations like GDPR \citep{regulation2018general} have made data collection increasingly difficult and expensive.
This looming data scarcity suggests that scaling laws may hit a wall \citep{hardy2024wall}, potentially impeding further AI advancement.

\textbf{Computational resources} has been the primary \textit{engine} powering AI development. Throughout AI history, major breakthroughs have been closely tied to advances in computing power, from early models requiring single CPUs (with peak performance of 1-2 GFLOPS) to modern GPU clusters. The computational demands have grown exponentially - from BERT-Large's training requiring 64 TPU v3 chips (providing 420 TFLOPS) \citep{devlin2018bert} to GPT-3's training on 10,000 V100 GPUs (reaching 28,000 TFLOPS) \citep{brown2020language}, while training GPT-4 reportedly required over 25,000 NVIDIA A100 GPUs (delivering a staggering 400,000 TFLOPS) \citep{openai2023gpt4}. More recent models like Grok 3 push these requirements even further \citep{xai2025grok3}. However, we are approaching physical limits in single-chip performance as Moore's Law slows down \citep{thompson2021deep}. While massive computing clusters can compensate for individual chip limitations, maintaining such infrastructure incurs astronomical costs - estimated at over \$100M for training GPT-4 \citep{sharir2020cost} - and poses significant environmental concerns due to their enormous energy consumption, with each training run emitting as much CO$_2$ as 500 cars driven for a year \citep{schwartz2020green}. Moreover, this level of computing power has become concentrated among a few tech giants, creating a monopolistic landscape that effectively excludes smaller companies and academic institutions from participating in foundational AI research \citep{hagiu2025artificial}. This centralization of computing resources presents a significant barrier to innovation and democratization in AI development \citep{anderljung2023frontier}.

In this paper, we propose that leveraging massive distributed edge devices offers a promising solution to overcome both data and computing barriers in AI development. 
Our analysis (using smartphone as an example) reveals two compelling opportunities: 
First, edge data generated from smartphones for past 5 years are projected to reach 33.1 EB, offering fresh, diverse, and contextually rich training samples. 
Second, the collective computing power of edge devices - with smartphones delivering 9,278 EFLOPS for past 5 years - demonstrates the feasibility of distributed model training, as training state-of-the-art models like DeepSeek-v3 would require only about 60,723 users with edge devices working (\textit{ideally}) in parallel to match its current training setup. 
Based on these insights, \textbf{we argue that leveraging these massive distributed edge devices can break barriers of data and computing wall, and everyone can participate in training large models with small edge devices.} 
To support this position, we first analyze the critical challenges of large language models, examining both data bottlenecks (\S\ref{subsec:data_exhaustion}) and computational monopolization (\S\ref{subsec:compute_monopoly}). 
We then explore the hidden potential of massive edge devices, investigating their vast untapped distributed data resources (\S\ref{subsec:edge_data}) and computational capabilities (\S\ref{subsec:edge_compute}). 
Building on these insights, we investigate technical approaches for overcoming large model challenges through distributed computing architectures (\S\ref{sec:technical_advancements}): small language models at edges (\S\ref{subsec:small_language_models}), collaborative inference (\S\ref{subsec:collaborative_inference}), and collaborative training (\S\ref{subsec:collaborative_training}).
We then identify two critical open challenges: heterogeneous device model fusion and heterogeneous device compute sharing (\S\ref{sec:open_problem}). 
Finally, we discuss the societal impact like AI democratization, incentive mechanisms, and environmental benefits of this paradigm shift (\S\ref{sec:impacts}).

\section{Scaling at Risk: Challenges of Data and Computing Power}
\subsection{The Ceiling of Public Data}
\label{subsec:data_exhaustion}
\paragraph{Public data for pretraining is exhausting.} The rapid advancement of large language models has created an insatiable appetite for training data. Scaling laws establish that model performance improves predictably with data quantity—a relationship that demands exponentially growing datasets \cite{hoffmann2022training}. A canonical example is GPT-3, trained on {300 billion tokens} spanning books, web content, and programming code \cite{brown2020language}. Current projections suggest dataset sizes grow at 0.38 orders of magnitude (2.4$\times$) annually \cite{villalobos2022trends}, implying models will require {three orders of magnitude more data} within a decade. \looseness-1

Despite the internet's vast textual resources, the total stock of high-quality human-generated text remains bounded. Recent estimates place this limit at approximately {$4\times 10^{14}$ tokens} \cite{villalobos2022trends}. \cite{villalobos2024will} argues that current consumption patterns suggest exhaustion of public text data by 2028, potentially accelerated to 2026 through excessive data reuse during training (a practice termed {overtraining}). 
Therefore, the finite nature of publicly available human-generated text data is expected to become a major bottleneck for LLM scaling within the next decade. Despite the current large scale of public data, the risk of data exhaustion is rapidly approaching as data demand continues to grow \cite{sevilla2022compute}.

\paragraph{Synthetic data has potential but faces challenges.}
Faced with the threat of data exhaustion, researchers have proposed various solutions, among which synthetic data generation is considered one of the most promising approaches. 
By leveraging LLMs to produce their own training data, researchers envision {self-sustaining data ecosystems}. Early successes in constrained domains like mathematics and code generation, where automated verification ensures quality, demonstrate potential \cite{liu2023tinygsm}.  Recent work \cite{chen2024diversity} demonstrated that diverse synthetic data enhances the performance of LLMs during both pre-training and fine-tuning.

The adoption of synthetic data faces three fundamental challenges. First, \textit{model collapse} occurs when models iteratively train on their own outputs, causing gradual divergence from original data distributions. This recursive process amplifies biases and reduces output diversity, ultimately degrading model performance across generations \cite{shumailov2023curse,dohmatob2024strong}. 
Second, \textit{synthetic data quality} remains inherently unverifiable in open-domain contexts. While formal domains like mathematics allow algorithmic validation, natural language lacks objective evaluation standards. The absence of ground-truth verification creates self-referential quality assessments, compromising reliability \cite{alemohammad2023self,wenger2024ai}.
Finally, synthetic data struggles to replicate human \textit{linguistic diversity}. Current methods disproportionately replicate dominant language patterns while underrepresenting cultural nuances and low-frequency expressions. This homogeneity limits their utility for training robust general-purpose models \cite{fan2024scaling}.

These persistent challenges underscore that synthetic data alone cannot sustainably address the looming data scarcity crisis, compelling the research community to seek complementary strategies that transcend conventional data acquisition paradigms.

\subsection{The Monopoly of Computing Resources}\label{subsec:compute_monopoly}
\paragraph{A few AI giants dominate the computing power.}
The AI computing landscape is dominated by a few major tech giants like OpenAI, Google, Microsoft, and Meta, which control powerful hardware such as GPUs and TPUs. This monopolization creates a significant barrier for smaller AI startups and research institutions, who struggle to access such advanced resources. Additionally, these companies control proprietary AI models, datasets, and software frameworks that require immense computing power, further widening the gap between the giants and smaller players. As a result, high-performance computing resources remain increasingly inaccessible to anyone outside these dominant entities.

\begin{wrapfigure}{r}{0.5\linewidth}
    \vspace{-12pt} %
    \centering
    \includegraphics[width=\linewidth]{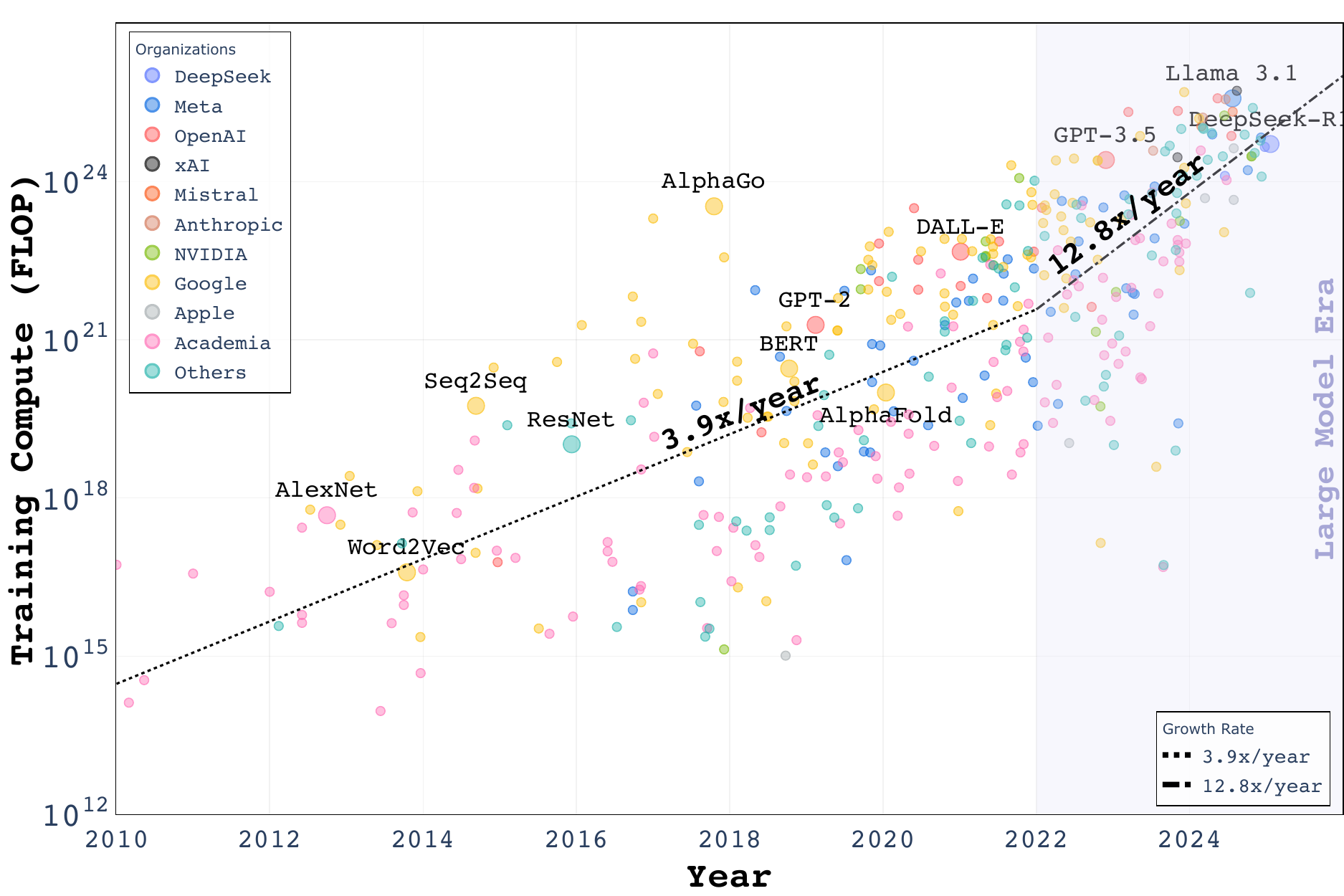}
    \vspace{-15pt} %
    \caption{Trend of Computational Demand for Model Training. (Data source:~\cite{epoch2023trendsinmachinelearninghardware}).}
    \label{fig:model_comput_trend}
    \vspace{-12pt} %
\end{wrapfigure}

\paragraph{Computational demand is growing exponentially.}
As large-scale AI models like GPT-4~\cite{openai2023gpt4}, Llama 3~\cite{meta2024llama3.1}, and DeepSeek-V3~\cite{liu2024deepseek} surpass the trillion-parameter scale, the global AI landscape faces severe computational efficiency challenges. As shown in Figure~\ref{fig:model_comput_trend}, since the deep learning revolution in 2010, AI training demands have grown at a super-exponential rate of 3.9$\times$ per year—an acceleration that intensified with the adoption of the transformer architecture as the industry standard \cite{vaswani2017attention}. With the advent of the era of large language models in 2022, the demand for computing power has surged even further, reaching an unprecedented growth rate of 12.8$\times$ per year. This marks a transformative shift in AI computation, where the need for computing power is expanding at an unprecedented pace, pushing the limits of existing hardware and infrastructure. \looseness-1

\paragraph{Moore's Law is slowing down.}
Moore's Law, which has driven the growth in computing power for decades, is slowing down as we approach the physical limits of silicon-based chip technology~\cite{kressel2023end}. The difficulty in shrinking transistors has led to diminishing returns in computational performance. As a result, the AI industry is relying more on specialized hardware like GPUs, TPUs, and custom chips to meet growing demands. However, this shift has made high-performance hardware even more expensive and exclusive, further intensifying the gap between organizations with the resources to develop advanced AI models and those without.

\paragraph{Infrastructure capacity is a constraint.}
The rapid expansion of AI model scales and the surge in computational demand are facing dual constraints in global computing infrastructure. On one hand, bottlenecks in advanced semiconductor manufacturing severely limit the expansion rate of AI data centers. The foundry capacity for wafers at 5nm and below—such as those produced by TSMC—has already been fully booked by leading technology companies until 2026~\cite{benzinga2024tsmc}. Moreover, the construction of new wafer fabs involves long lead times and is further constrained by the global supply chain shortages of critical equipment, such as lithography machines.
On the other hand, the exponential increase in chip deployment within individual AI clusters is putting immense pressure on the already limited semiconductor manufacturing capacity, pushing the industry toward its production ceiling~\cite{scaleflux2024ai}.
These factors have significantly hindered the continuous expansion of computing power, making it increasingly difficult to scale AI infrastructure sustainably.

\section{Scaling Beyond Limits: Opportunities from Edge Devices}
\label{sec:opportunities}

\subsection{Massive Data from Edges}
\label{subsec:edge_data}

As discussed in §~\ref{subsec:data_exhaustion}, edge data represents a crucial alternative to synthetic data in addressing the challenge of data exhaustion. Edge data refers to the data generated by edge devices at or near the source of data generation, which typically remains private and localized rather than being publicly accessible. Edge devices encompass a wide range of equipment including Internet of Things (IoTs) sensors, smartphones, wearables, industrial controllers, and other smart devices that process data at the network edge. 
Data generated at edges offers unique advantages in both data volume and data quality. \looseness-1

\begin{figure*}[htbp]
    \centering
    \begin{minipage}[t]{0.49\linewidth}
        \centering
        \includegraphics[width=\linewidth]{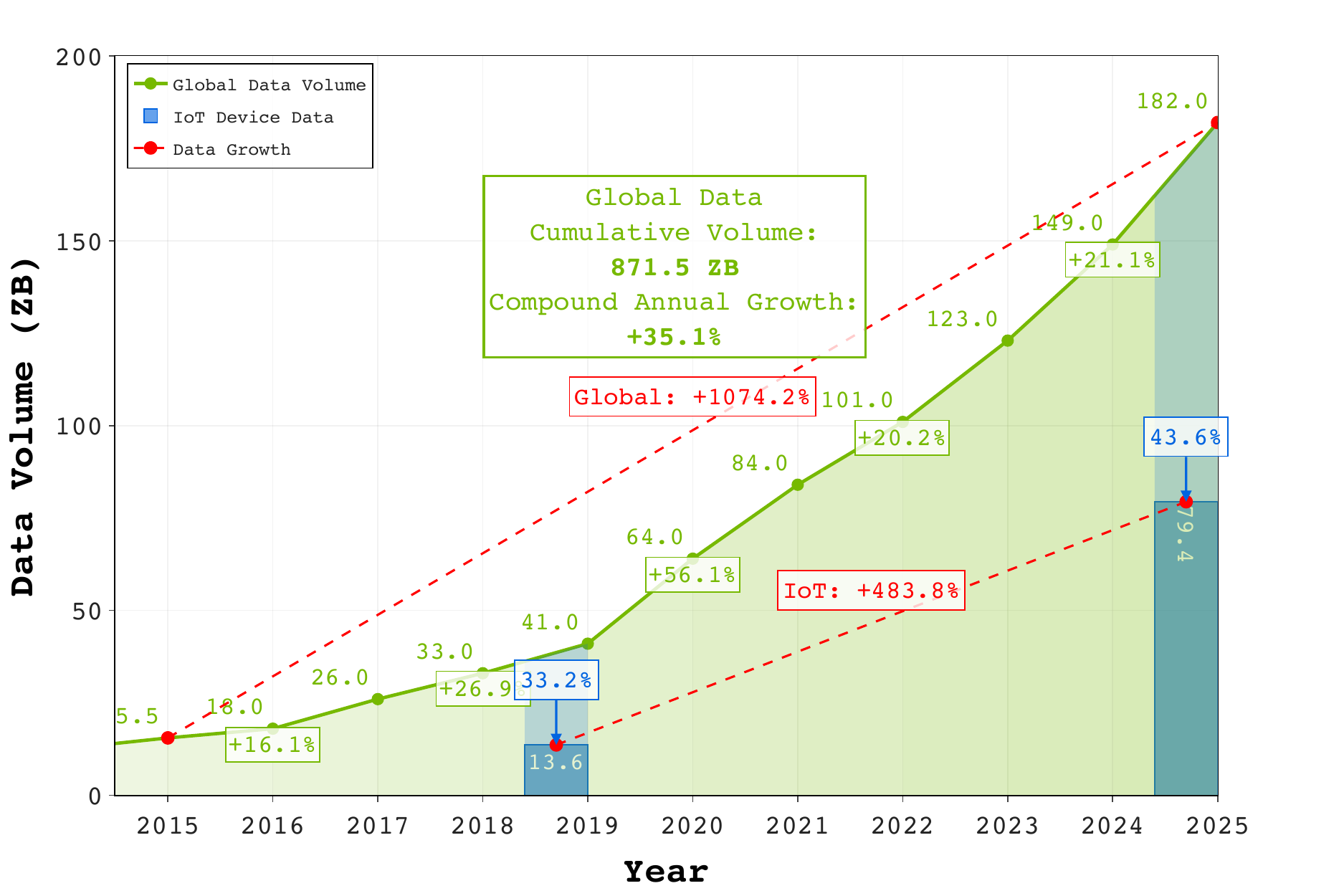}
        \caption{Global data volume from 2014 to 2025 and IoT device data volume in 2015 and 2025. (Data sources: Global data volume from \cite{statista_global_2023}; IoT device data volume from \cite{statista_iot_2023}.)}
    \label{fig:global_data}
    \end{minipage}
    \hfill
    \begin{minipage}[t]{0.49\linewidth}
        \centering
        \includegraphics[width=\linewidth]{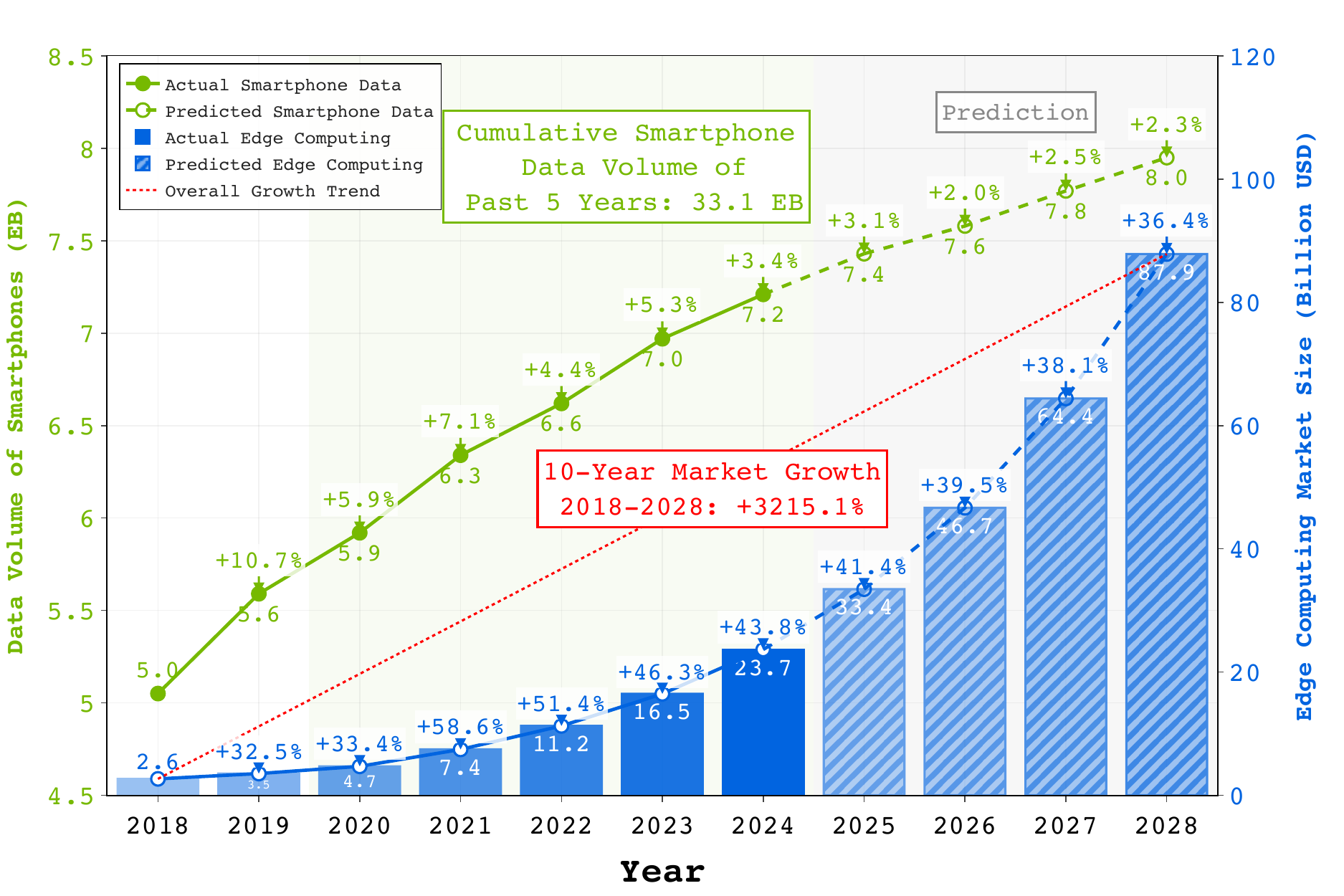}

        \caption{Smartphone data volume with edge computing market size (right) from 2018 to 2028. (Data sources: Edge computing market \cite{grandview_edge_2023}; Smartphone data volume from \cite{bankmycell_smartphone_2023}.)}
    \label{fig:smartphone}
    \end{minipage}
    \vspace{-10pt}
\end{figure*}

\paragraph{Edge-generated data is explosively growing.} 
According to the statistical data from \cite{statista_global_2023,statista_iot_2023} (as illustrated in Figure~\ref{fig:global_data}), 
the global data volume is projected to reach 182 ZB by 2025~\cite{statista_global_2023}, where the data generated by IoT devices is anticipated to increase from 13.6 ZB in 2019 to 79.4 ZB in 2025~\cite{statista_iot_2023}, elevating its share of the global data volume from 33.2\% to 43.6\%, showing a particularly pronounced growth in edge-generated data. Over the period from 2015 to 2025, the global data volume exhibited a compound annual growth rate (CAGR) of 35.1\%, resulting in an overall increase of 1074.2\% and a cumulative total of 871.5 ZB. IoT device data experienced a growth of 483.8\% from 2019 to 2025. This trend underscores the increasingly central role of edge-generated data in the global data ecosystem.
Beyond IoT devices, smartphones, as a critical source of edge-side data, are also contributing to the steady rise in data volume. As depicted in Figure~\ref{fig:smartphone}, the estimated smartphone data volume is projected to grow from 5 EB in 2018 to 8 EB by 2028
\footnote{These numbers are estimated based on an average user data generation of 1 GB per device. For detailed estimation methodology, refer to Appendix \ref{app:smartphone_ethod}.}. 
This exponential growth is closely aligned with the rapid expansion of the edge computing market, which is forecasted to surge from \$5.5 billion in 2019 to \$87.9 billion by 2028, representing a remarkable growth rate of 3215.1\%. The burgeoning edge computing market has further catalyzed the generation and processing of edge-side data, reinforcing its significance in the broader data landscape 
\footnote{\cite{idc_seagate_dataage_2019,seagate_rethinkdata_2020} provide a more comprehensive overview of the global data volume, but we cannot access the statistics data. 
We appreciate any suggestions for better statistical data sources.
}.

\paragraph{Edge-generated data has distinctive advantages.} 
Beyond its impressive quantity, edge data possesses several distinctive characteristics that make it particularly valuable for model training.
First, edge data provides enhanced \textit{privacy} characteristics. Since edge data typically remains local to devices and does not need to be centrally stored, it allows for more privacy-preserving approaches to data utilization. This local-first nature enables compliance with increasingly strict data privacy regulations while still allowing the data to contribute to model training.
Second, edge data exhibits superior \textit{diversity} across multiple dimensions. It encompasses a wide variety of data types from IoT devices, mobile interactions, and personal devices, covering different domains, languages, and user behaviors. This natural diversity provides richer training signals compared to curated public datasets \cite{nayak2024review}. 
Third, edge data demonstrates strong \textit{real-time} capability. Unlike public datasets, which are often updated infrequently, edge devices continuously generate fresh data with low latency \cite{cavliwireless_edgecomputing}, offering more up-to-date and relevant training samples.
Despite these advantages, edge-generated data can present challenges such as low signal-to-noise ratios, unclean, or potentially harmful content. However, recent advancements in data quality assessment methods (e.g., data prospector \cite{ni2024small}), have emerged to identify and filter high-quality data from edge sources, ensuring that only the most reliable and valuable data is selected for model training.

In conclusion, edge data with its explosive growth and distinctive characteristics, is a valuable resource for model pretraining. Its diversity, real-time nature, personalization, and rich context make it an ideal foundation for developing robust and adaptable large-scale models, enhancing their ability to serve real-world applications effectively.

\begin{insights}
    \textbf{Insight:} The smartphone data volume of the past 5 years (before 2025) is projected to reach approximately 33.1 EB, with unique advantages in privacy, diversity, and real-time context, demonstrating the massive data potential of edge for AI model training.
\end{insights}

\subsection{Massive Computing Power from Edges}\label{subsec:edge_compute}

\begin{figure*}[htbp]
    \centering
    \begin{minipage}[t]{0.49\linewidth}
        \centering
        \includegraphics[width=\linewidth]{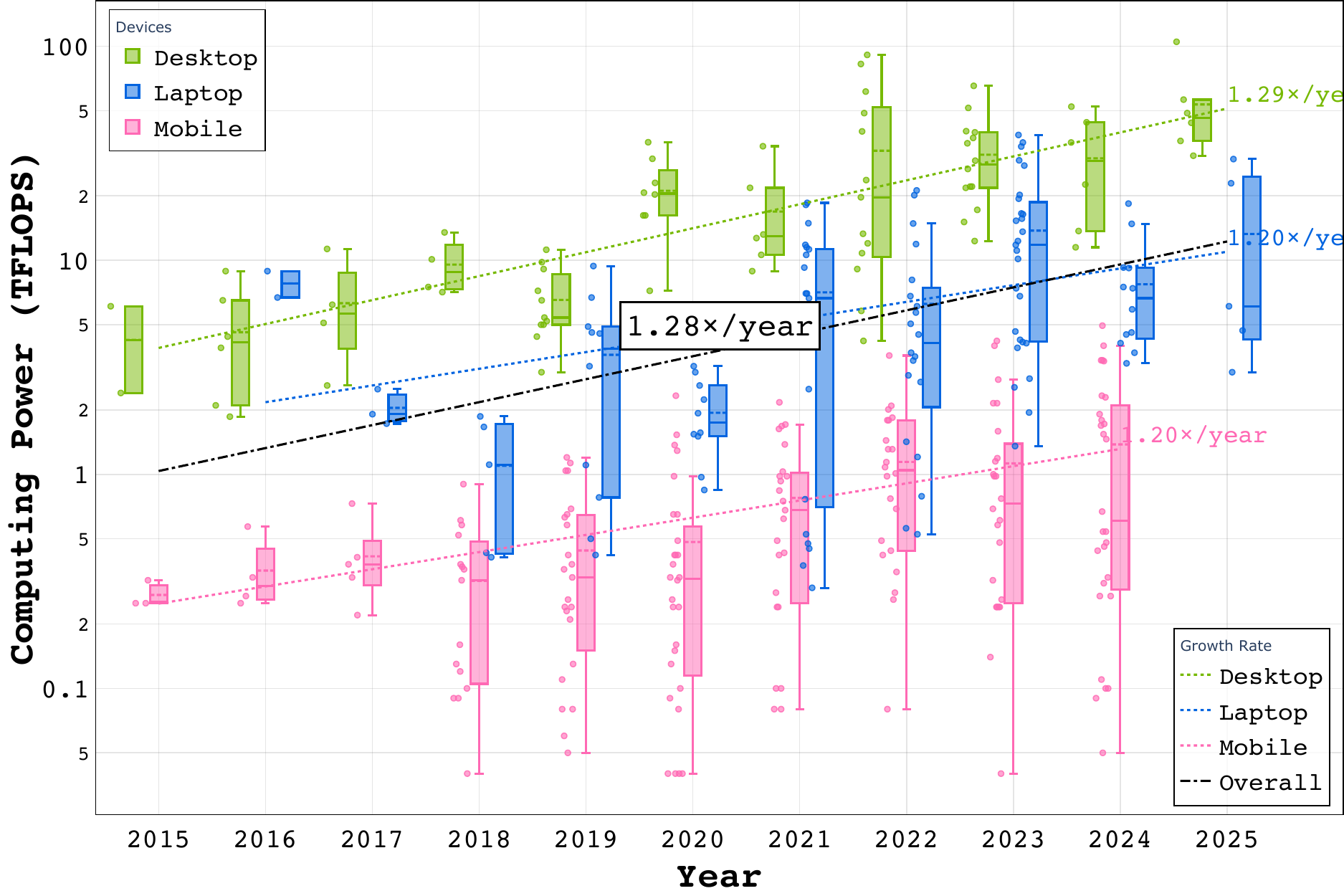}
        \caption{Edge Computing Power Evolution Trend. (Data source: \cite{nanoreview2025}).}
        \label{fig:edge_trend}
    \end{minipage}
    \hfill
    \begin{minipage}[t]{0.49\linewidth}
        \centering
        \includegraphics[width=\linewidth]{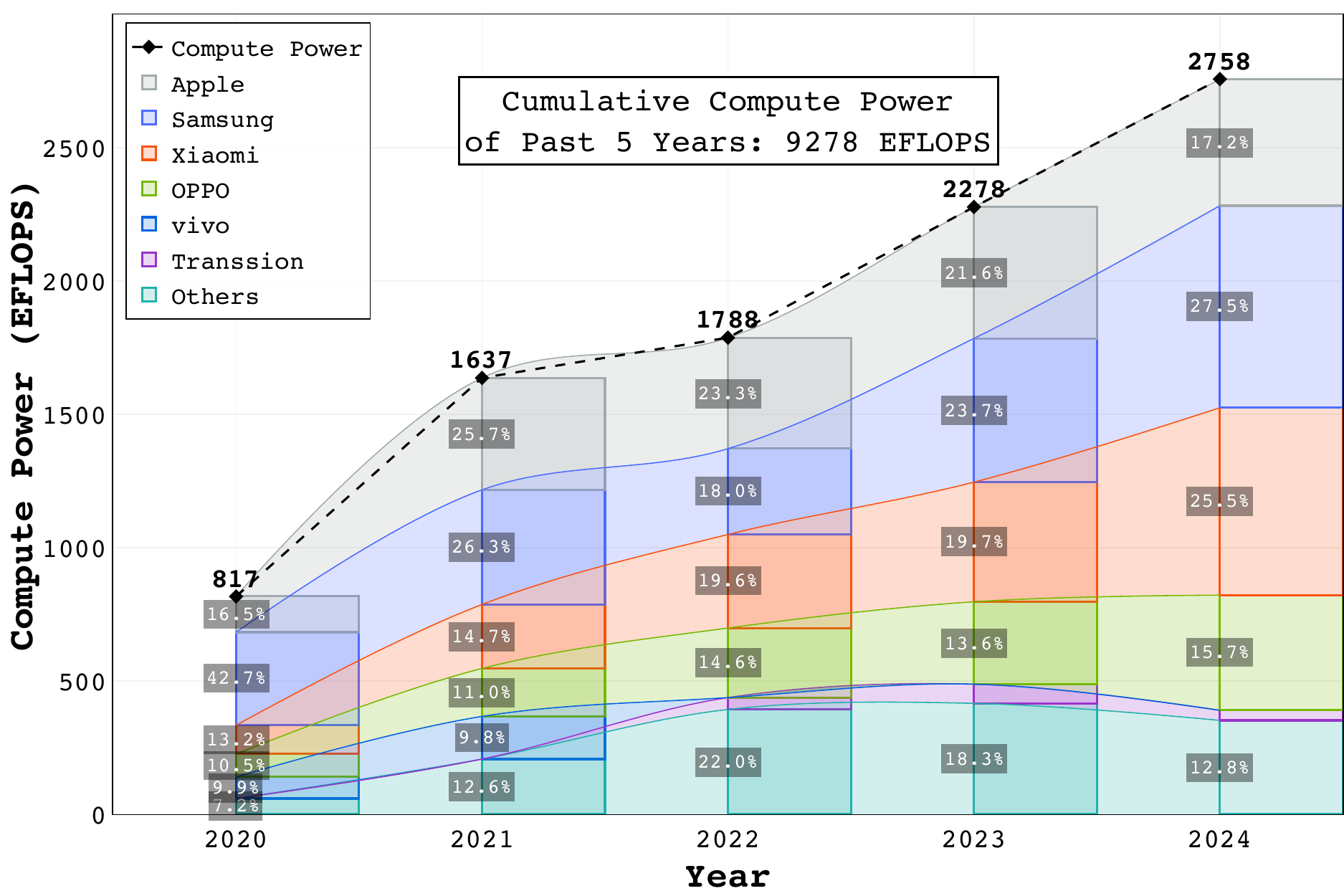}
        \caption{Smartphone Market Share and Computing Power Trends. (Data source: \cite{canalys2025}).}
        \label{fig:smartphone_trend}
    \end{minipage}
    \vspace{-10pt}
\end{figure*}

\paragraph{Edge computing power is growing rapidly.}
In recent years, edge computing has experienced explosive growth in computing power, driving smart devices to evolve from single-function tools into multimodal perception and decision-making centers. For instance, as shown in Figure~\ref{fig:edge_trend}, flagship smartphones such as the iPhone 16 series, equipped with 3nm process chips, have achieved computing power exceeding 2 TFLOPS~\cite{apple2024iphone16pro}, enabling local execution of complex AI tasks like real-time image enhancement and multilingual speech translation. 
Notably, the computing power of an individual smartphone has potentially surpassed that of laptops in the same period.
The breakthroughs are even more pronounced in the desktop sector,
achieving an annual computing power growth rate of 1.29$\times$/year, surpassing that of smartphones and laptops (both 1.20$\times$/year). 
These three types of devices form a differentiated growth hierarchy, collectively driving edge computing's overall computing power to expand at an annual average rate of 1.28$\times$/year.
This growth is fueled by three key technological drivers: advanced manufacturing processes (3nm technology increases transistor density by 60\%~\cite{apple2024iphone16pro}), dedicated architectures (modern smartphone SoCs integrate NPUs for AI acceleration), and scenario-driven innovation (e.g., autonomous driving demands end-to-end latency of less than 100ms~\cite{nvidia2023}).

\paragraph{Edge computing has potential for LLM training.}
We analyze the performance of smartphone chips, representing typical edge devices, and estimated their overall computing power.
To ensure our estimation is as accurate as possible, we based our calculations on the market share data from \cite{canalys2025}.
We then estimated the total computing power of newly produced mobile devices by averaging the chip performance of each brand.
From 2020 to 2024, smartphone chip performance has seen significant improvements, with peak computing power increasing from 1.53 TFLOPS to 4.95 TFLOPS, and average computing power rising from 0.48 TFLOPS to 1.38 TFLOPS. Meanwhile, the overall computing power of mobile devices has grown from 817 EFLOPS in 2020 to 2,758 EFLOPS in 2024, and totally 9278 EFLOPS for past 5 years. This trend highlights the rapid expansion of edge computing power, which is not only essential for AI applications but also holds the potential for training complex AI models.
For instance, training the DeepSeek-v3~\cite{liu2024deepseek} model utilizes 2048 H100 GPUs, each providing a peak FP32 performance of 59.30 TFLOPS, resulting in a total computational capacity of 121,446.4 TFLOPS. If this workload were distributed across edge devices with a peak performance of 2 TFLOPS (e.g., mobile chips like the iPhone 16 series), approximately 60,723 users with edge devices working (\textit{ideally}) in parallel would be required to match the computational capacity. \looseness-1

\begin{insights}
    \textbf{Insight:} 
    The smartphone computing power of the past 5 years (before 2025) is projected to reach approximately 9278 EFLOPS, with individual flagship devices now achieving over 2 TFLOPS performance. The combined parallel computing power of approximately 30 iPhone devices (with A18 chips) can match the computational capacity of a professional AI training GPU (H100 with 59.30 TFLOPS). 
\end{insights}

However, current smartphone chips are primarily optimized only for inference efficiency rather than training capabilities. 
\textit{We advocate for and predict a future trajectory of edge computing where smartphone chip designs will increasingly prioritize and optimize on-device model training capabilities.
}
As computational power grows and distributed algorithms develop, we expect a paradigm shift enabling collaborative model training across networks of edge devices. This evolution positions the edge computing ecosystem as a critical catalyst for democratizing AI development and driving the next wave of innovations in the field.

\section{Technical Advancements}\label{sec:technical_advancements}

\subsection{Small Language Models at Edges}\label{subsec:small_language_models}

The first move of AI to Edge is to deploy small language models (SLMs) to edge devices \cite{lu2024small,wang2024comprehensive,van2024survey}. This trend is driven by the growing demand for AI applications that can run directly on edge devices, motivated by needs for privacy, offline usage, and real-time processing without cloud dependence. However, edge devices have limited memory, computation, and energy resources, requiring more efficient and compact models.

\paragraph{SLMs leverage innovative architectures for efficient edge deployment.}
The classic Transformer architecture \cite{vaswani2017attention} uses self-attention mechanisms for effective sequence modeling but faces quadratic complexity challenges, with models like TinyBERT \cite{jiao2020tinybert} (14.5M parameters) and ALBERT \cite{lan2020albert} (12M parameters) demonstrating its effectiveness at small scales. Mamba \cite{gu2023mamba}, based on state space models, achieves linear complexity and faster inference by utilizing only the previous hidden state, as demonstrated by Zamba2-2.7B \cite{glorioso2024zamba2} which achieves twice the speed and 27\% reduced memory overhead compared to traditional models. Hymba \cite{dong2024hymba} combines both approaches by integrating attention and SSM heads within the same layer for parallel processing, with its 1.5B variant trained on DCLM-Baseline-1.0 and SmolLM-Corpus achieving 11.67 times cache size reduction while outperforming Llama-3.2-3B. The xLSTM architecture \cite{beck2024xlstm} modernizes LSTM with exponential gates and matrix memory cells, with models ranging from 125M to 1.3B parameters trained on 300 billion tokens from SlimPajama \cite{cerebras2023slimpajama}, consistently outperforming comparable RWKV-4 \cite{peng-etal-2023-rwkv}, Llama \cite{touvron2023llama3}, and Mamba models across various tasks in the PALOMA benchmark \cite{magnusson2024paloma}. These architectural innovations demonstrate the potential for efficient and powerful language models that can run effectively on edge devices.

\paragraph{SLMs can be constructed through diverse methodological approaches.}
The construction of efficient SLMs relies on a comprehensive suite of techniques, each with specific performance trade-offs. 
For training SLMs from scratch, optimized MLM approaches \cite{devlin2019bert} with increased masking ratio (25\% vs traditional 15\%) demonstrate 2-3\% performance improvements for models under 3B parameters.
When deriving SLMs from existing LLMs, knowledge distillation has proven particularly effective, with response-based distillation \cite{sanh2019distilbert} reducing model size by 40\% while maintaining 95\% of the original performance.
In architecture optimization, the Mixture of Experts approach \cite{fedus2022switch} enables 65\% parameter reduction while potentially improving performance in specific tasks. Domain specialization has shown remarkable results, particularly in the medical field where 3B parameter models achieve 92\% accuracy, outperforming 175B models (89\%) \cite{fu2023specializing}.
The combination of these techniques yields impressive results - a notable example is a 770M parameter model from \cite{hsieh2023distilling} that combines distillation, quantization, and domain specialization to achieve 95\% of the performance of a 540B model on specific tasks while requiring less than 0.15\% of the computational resources. Most successful SLMs achieve their efficiency by combining multiple techniques, typically starting with knowledge distillation, applying compression methods, and finishing with domain-specific fine-tuning. 
\cite{wang2024comprehensive} has provided a comprehensive survey of SLMs, and we summarize the architecture innovations and training methods in Appendix \ref{app:slm_architectures_training}. \looseness-1

Despite remarkable progress in deploying compressed models to edge devices, the current landscape remains largely confined to individual devices operating in isolation, failing to leverage the massive distributed computing power that could be achieved through collaborative training across edge devices. This represents a significant missed opportunity, as the collective computing resources of billions of edge devices worldwide remain untapped, while individual devices struggle with the computational demands of modern AI applications.

\subsection{Collaborative Inference at Edges}\label{subsec:collaborative_inference}
The emergence of collaborative inference at the edge represents a significant shift in AI infrastructure, enabling more accessible and cost-effective AI solutions compared to traditional data centers which often present barriers in terms of cost, energy consumption, and accessibility. Recent frameworks like Exo \cite{exo} enable users to create AI clusters using everyday devices such as phones, tablets, and computers through peer-to-peer architecture, effectively unifying their computational resources.

Several approaches advance this paradigm: Neurosurgeon \cite{kang2017neurosurgeon} introduces a lightweight scheduler that partitions DNN computations between devices and datacenters; MoE$^2$ \cite{jin2025moe2} optimizes LLM inference under energy and latency constraints; Edgent \cite{li2018edgent} enables low-latency edge intelligence through adaptive partitioning and early-exit mechanisms; and Galaxy \cite{ye2024galaxy} leverages hybrid model parallelism to efficiently execute Transformer models across edge devices. Despite these advances, the current landscape has yet to fully capitalize on the potential of distributed data resources at the edge. The next frontier lies in transforming these devices from mere inference endpoints into active training participants, representing a significant opportunity for distributed AI development. \looseness-1

\subsection{Feasibility of On-Device Training}

Recent advancements in optimization techniques have significantly reduced the memory and computational requirements for training machine learning models, making on-device training increasingly feasible, even on resource-constrained edge devices.
For instance, Lin et al. \cite{lin2022ondevice} demonstrated training neural networks on microcontroller units with only 256KB of RAM by employing an algorithm-system co-design framework. 
Similarly, Cai et al. \cite{cai2020tinytl} introduced a tiny transfer learning approach that freezes most parameters and only trains a small subset, allowing effective fine-tuning with minimal memory requirements. Qiu et al. \cite{qiu2022zerofl} proposed ZeroFL, a framework that relies on highly sparse operations to accelerate on-device training in federated learning settings, enabling efficient model training on edge devices with up to 95\% sparsity.
Recent work by Sugiura and Matsutani \cite{sugiura2025elasticzo} further advanced this field with ElasticZO, which combines zeroth-order and first-order optimization to achieve a better trade-off between accuracy and training cost. Their ElasticZO-INT8 variant achieves integer arithmetic-only training, further reducing memory usage and training time by approximately 1.5x without compromising accuracy.
These advancements suggest that on-device training is no longer limited to high-end devices with abundant resources. Even small embedded systems with memory capacities measured in kilobytes rather than gigabytes can participate in model training.\looseness-1

\subsection{Collaborative Training at Edges}\label{subsec:collaborative_training}

To harness the potential of vast amounts of data distributed across numerous devices, we envision a future where everyone can participate in training large-scale models. Federated learning emerges as a paradigm for distributed collaborative training that makes this vision possible.

\textbf{Federated learning} \cite{mcmahan2017communication} is a practical paradigm that enables collaborative model training while preserving data privacy. Instead of collecting raw data from edge devices, which may violate privacy regulations like GDPR \cite{regulation2018general}, this approach distributes the training process across multiple devices. Each device trains on its local data and only shares model updates with the central server. This approach can effectively utilizes both computational and data resources available at edges.

\textbf{Federated LLMs for fine-tuning}
has emerged as a critical direction in recent research of large language models, addressing the challenges of privacy preservation and resource constraints. FedDAT \cite{chen2024feddat} introduces a framework for foundation-model fine-tuning under multi-modal heterogeneous federated settings, while FedFM \cite{yang2025federated} tackles the critical challenges of system and statistical heterogeneity in federated learning through adaptive optimization across diverse client devices.
To address the computational constraints of edge devices, FedPETuning \cite{zhang2023fedpetuning} employs parameter-efficient fine-tuning techniques, significantly reducing the memory and computation requirements while maintaining model performance. This approach enables even resource-constrained devices to participate in the fine-tuning process. Similarly, \cite{chen2024feddat} bridges the gap between federated learning and foundation models by introducing novel techniques for efficient knowledge transfer and model adaptation in multi-modal heterogeneous federated learning settings.
In domain-specific applications, FedMatch \cite{chen2021fedmatch} demonstrates the effectiveness of federated learning for question-answering tasks, showing that models can be fine-tuned on sensitive domain-specific data while preserving privacy. These advancements are supported by open-source frameworks like Flower \cite{Flower} and FATE-LLM \cite{fan2023fate}, which provide robust platforms for implementing federated fine-tuning of large language models.

\textbf{Federated LLMs for pretraining} 
have opened up exciting new possibilities for large language model training. Rather than relying on traditional data center approaches, researchers have developed innovative geographically distributed frameworks that enable collaborative training across many devices.
Notably, Prime Intellect \cite{OpenDiLoCo} has launched the first decentralized training project for a 10 billion parameter model, named INTELLECT-1, which utilizes the OpenDiLoCo framework to significantly reduce communication costs between nodes. This innovative approach allows for dynamic management of computational resources across multiple locations, achieving an impressive 83\% overall computational utilization while collaborating with up to 112 H100 GPUs across five countries and three continents. The model not only enhances parameter efficiency by 25 times but also demonstrates robust performance in various benchmark tests.
In parallel, the Flower Lab \cite{Flower} has introduced FlowerLLM, which successfully trained a 1.3 billion parameter large language model (LLM) using novel federated learning methods. 
Additionally, it has also developed Photon \cite{sani2024photon}, an open-source framework that provides flexible configurations for training models of different sizes, making federated LLM training more accessible to a broader range of participants and computational resources. \looseness-1

These frameworks underscore a shift towards decentralized AI inference and training (summarized in Appendix~\ref{app:distributed_collaborative_frameworks}), enabling researchers worldwide to contribute to advanced model development without the constraints of centralized resource control, thus paving the way for a more collaborative and inclusive AI landscape. \looseness-1

\section{An Open Problem: How to Train LLMs with Small Edge Devices?}\label{sec:open_problem}

A fundamental limitation of traditional federated learning lies in its requirement for each participant to maintain and train a complete model locally. This assumption becomes particularly problematic in the context of large language models, where the computational and memory requirements far exceed the capabilities of most individual participants. For instance, in sensitive domains like healthcare~\cite{bolton2024biomedlm}, multiple hospitals may wish to collaboratively train a medical language model to leverage their collective data while maintaining data privacy. However, traditional federated learning mandates that each participating hospital possess sufficient computational resources to train the complete model locally. Despite these institutions' valuable data contributions and their motivation to enhance model capabilities through collaborative training, many hospitals lack the necessary infrastructure to participate effectively. This resource constraint significantly limits the potential for collaborative model training in critical domains where data privacy is paramount but computational resources are unevenly distributed \cite{kairouz2021advances}. 
\begin{figure}
    \centering
    \includegraphics[width=0.8\linewidth]{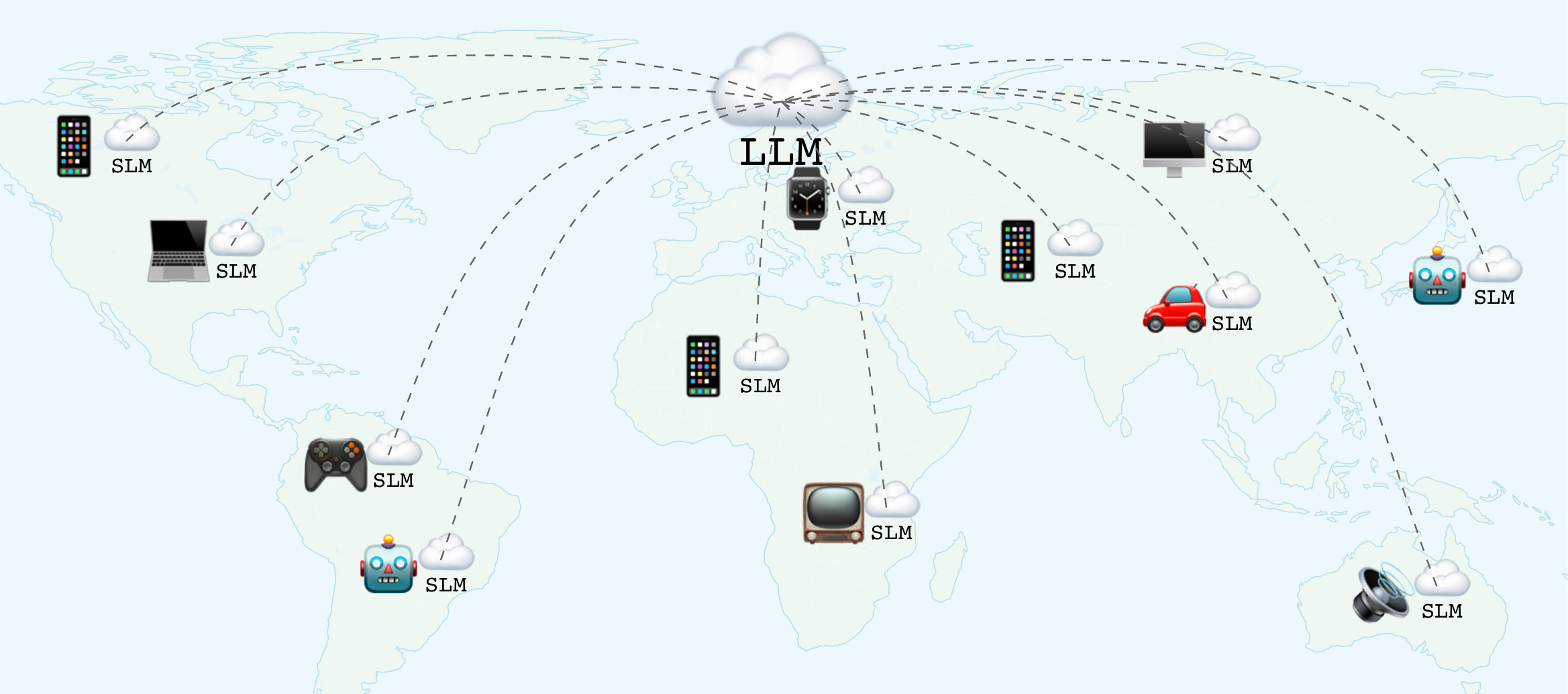}
    \caption{Train Large Language Models with Small Edge Devices}
    \vspace{-15pt}
    \label{fig:fedllm}
\end{figure}
While some federated learning approaches allow training partial model parameters \cite{horvath2021fjord,alam2022fedrolex}, the enormous disparity between large language models and what edge devices can train—often orders of magnitude smaller due to inherently constrained resources—remains too vast to be effectively bridged by current FL frameworks.
Therefore, we need to develop a new federated learning paradigm that enables participants to collaboratively train a large language model even under extremely limited resources (as shown in Figure~\ref{fig:fedllm}).
\textbf{It is still an open problem to train large language models with small edge devices.} Therefore, we encourage the research community to develop novel distributed collaborative computing methods in two key directions:

\subsection{Heterogeneous Device Model Fusion: from Small to Large}\label{subsec:model_fusion}

The first direction addresses the fundamental challenge of model size disparity in federated learning. Modern large language models typically contain hundreds of billions of parameters, while edge devices have severely limited computational resources. This creates an enormous scale gap - the large target model may be hundreds or even thousands of times larger than what individual devices can handle. To bridge this gap, each edge device should run a small language model that matches its computational capacity. For example, while the central model may have 100 billion parameters, a resource-constrained mobile device might only handle a 100-million parameter model, representing a 1000x size difference. 
The key challenge then becomes how to effectively aggregate and fuse knowledge from these much smaller models into the large target model. 
We need novel techniques that can meaningfully combine insights from models operating at radically different scales while preserving the unique contributions of each small model. This requires fundamentally rethinking traditional model fusion approaches \citep{velasevic2023effects,azizan2019distributed} to handle such extreme parameter count disparities.

\subsection{Heterogeneous Device Compute Sharing: from Node to Cluster}\label{subsec:compute_sharing}

The second direction is to enable efficient compute resource sharing across heterogeneous devices by treating them as a unified compute cluster rather than independent nodes. Consider a smart home environment where multiple devices—smartphones, laptops, and desktop computers—could form a collaborative compute cluster. While each individual device has limited resources, their collective computing power could be substantial. For example, a laptop could handle intensive computational tasks, smartphones could manage coordination and lightweight processing, and desktop computers could contribute their onboard computing power. Meanwhile, other IoT devices such as smart speakers, security cameras, and vehicles could serve as data sources, providing valuable real-world inputs like voice commands, visual feeds, and environmental parameters. The language model would effectively run and train across this entire device cluster, leveraging both computing power and diverse training data from the environment.
This distributed execution requires new frameworks that can intelligently decompose and distribute model computations based on each device's capabilities and current load. The system must dynamically balance workloads - when the security cameras are idle at night, they could take on additional compute tasks, while during peak usage hours, the load could shift to other devices. 
This requires innovations in real-time resource allocation, task scheduling across heterogeneous hardware, and efficient inter-device communication protocols to ensure the collective computing power is optimally utilized \citep{zhao2024retrieval}.

\section{Conclusion}

In this position paper, we have argued that 
leveraging massive distributed edge devices can break barriers of data and computing wall, and everyone can participate in training large models with small edge devices.
Our comprehensive analysis demonstrated the vast untapped potential of edge resources, with smartphone data volume reaching approximately 33.1 EB and a combined computing power of around 9278 EFLOPS  in the past 5 years. 
These edge resources offer unique advantages in terms of data diversity, privacy, real-time context, and computing efficiency. 
This paradigm shift towards distributed training could democratize AI development and open an exciting new chapter in the scaling of foundation models.

\newpage
\appendix
\section{Impact Statements}\label{sec:impacts}

The shift from centralized to distributed training of large models, may introduce new technical and societal challenges and have the potential to fundamentally reshape the AI landscape.

\subsection{AI Monopoly and Democratization}\label{subsec:ai_monopoly_and_democratization}

The current AI landscape is characterized by significant concentration of power among a few tech giants, primarily due to their monopoly over massive computing resources and data centers \cite{bommasani2021opportunities}. This monopolistic trend has intensified with companies like OpenAI increasingly moving towards closed systems. 
While open-source alternatives like Llama \cite{touvron2023llama3}, Deepseek \cite{liu2024deepseek} and other community-driven models have made strides towards democratization by releasing model parameters and technical reports \cite{welsh2024democratising}, the gap in computational resources and data access between major AI companies and other players remains substantial and continues to widen. 
This disparity in resources allows tech giants to maintain their absolute dominance in determining the direction of AI development, raising concerns about AI democratization.

Edge device-based collaborative training presents a promising pathway to democratize AI development \cite{yao2022edge}.
By leveraging the collective computing power of millions of edge devices, this approach could effectively challenge the existing monopolistic structure \cite{dong2025beyond}. 
This democratization of AI training through edge devices could fundamentally reshape the structure of responsibilities and authorities.
If everyone can participate in training LLMs, the AI landscape could fundamentally change. Training decisions would shift from companies to communities, creating shared responsibility for model development \cite{zeng2023distributed}. Global participation would help models reflect diverse cultural perspectives, while allowing communities to adapt models for their local needs.
Furthermore, this decentralized approach could foster a more competitive and innovative AI ecosystem. When the barriers to entry for AI model training are lowered, we can expect to see a broader range of specialized models emerging (like \cite{song2025injecting,MedicalGPT,li2023large,yao2024lawyer,AlphaEvolve}), better suited to local needs and diverse use cases.

\subsection{Fairness and Incentive Mechanisms}\label{subsec:fairness_and_incentive_mechanisms}

The distributed training paradigm introduces new considerations for model fairness and bias mitigation \cite{kheya2024pursuit,li2020fair}. When training occurs across diverse edge devices, the resulting models can potentially better reflect the heterogeneous nature of user populations \cite{wang2020optimizing}. However, this approach also raises concerns about participation bias, where differences in device capabilities or user engagement could lead to underrepresentation of certain groups \cite{kairouz2021advances}. To address these challenges, researchers have proposed various fairness-aware federated learning algorithms \cite{mohri2019agnostic} that aim to ensure equitable model performance across different demographic groups and device types \cite{li2020fair}.

To sustain a distributed training ecosystem, effective incentive mechanisms are crucial for motivating user participation \cite{kang2019incentive}. Traditional approaches like computational resource sharing \cite{khan2020federated} and privacy-preserving reward systems have shown promise in encouraging user engagement. More innovative solutions include token-based reward systems \cite{wang2023incentive} and reputation mechanisms \cite{yang2019federated} that compensate users for their contributions while maintaining system integrity. These incentive structures not only encourage consistent participation but also help ensure the quality of contributed training data \cite{sharghivand2020edge}, creating a sustainable ecosystem for collaborative AI development.

\subsection{Carbon Footprint and Energy Efficiency}\label{subsec:carbon_footprint_and_energy_efficiency}
The shift from centralized to distributed training offers compelling environmental benefits \cite{guidi2024environmental}. Traditional data centers housing large language models face significant energy challenges\cite{sarkar2024carbon} - their high-performance GPUs require extensive cooling systems that consume 30-40\% of total energy \cite{cooling}. In contrast, FL distributes computation across edge devices like smartphones and tablets that operate at much lower temperatures and power levels, eliminating industrial cooling needs \cite{fl_vs_centralized}.

FL also dramatically reduces data transmission energy costs. While centralized approaches require raw data transfer from millions of devices, FL only transmits lightweight model updates, substantially decreasing network energy overhead \cite{fl_data_transmission}. The hardware efficiency gap is striking - edge devices like the NVIDIA Tegra X2 consume just 7.5W during training compared to 250W for data center GPUs \cite{hardware_power}, translating to major carbon footprint reductions, particularly for simpler models \cite{fl_iid}.
By reducing reliance on power-hungry data centers and leveraging existing consumer devices, FL enables more sustainable AI development through optimized energy efficiency and minimized infrastructure needs. This combination of reduced cooling requirements, efficient hardware utilization, and optimized data handling makes FL an environmentally responsible choice for the future of AI training \cite{fl_environmental}. As climate impact becomes increasingly critical, FL's sustainability advantages position it as a key technology for green AI development.

\section{Historical Development and Current Challenges}\label{sec:history}

\subsection{Data: the fuel of LLMs}

\paragraph{Early data-driven AI development}
As LLMs continue to achieve unprecedented success in artificial intelligence, understanding the role of data becomes increasingly crucial. From the early days of simple datasets to the modern era of massive data collections, data has consistently served as the lifeblood of AI, determining the upper bounds of model capabilities. The evolution of AI—marked by breakthroughs in computer vision, natural language processing, and beyond—can be traced back to the continuous expansion and refinement of data resources. 

In the early stages of AI, despite relatively small data scales, the importance of data was already evident. The MNIST dataset, for instance, serves as a notable example. With 60,000 training images and 10,000 test images, it provided a crucial foundation for neural network research, demonstrating the fundamental role of data in model training~\cite{lecun1998mnist}. As data scales expanded, the capabilities of deep learning models saw significant improvements. The emergence of ImageNet, which contains 14 million images across 21,000 synsets, revolutionized computer vision. This enabled deep learning models like AlexNet to learn complex visual features and achieve breakthrough progress in image recognition tasks, reducing error rates from 26.2\% to 15.3\% in the ILSVRC-2012 competition~\cite{deng2009imagenet,krizhevsky2012imagenet}. ImageNet's success stemmed not only from its scale but also from its high quality and diversity, laying the groundwork for subsequent large-scale data applications.

\paragraph{Era of massive data}
With the proliferation of the internet and advances in computing power, data scales have expanded dramatically, ushering AI into an era of massive data. GPT-3, for instance, was trained on 300 billion tokens, with a carefully curated mix of data sources: Common Crawl (60\%), books (16\%), Wikipedia (3\%), and other internet-based text (21\%)~\cite{brown2020language}. This massive dataset enabled GPT-3 to excel across various tasks, demonstrating the decisive role of data scale in model capabilities. Compared to early datasets like MNIST and ImageNet, GPT-3's data scale and quality reached unprecedented heights, not only advancing natural language processing but also opening new possibilities for AI generalization.

\paragraph{Quality and diversity matter}
Beyond scale, data quality and diversity are crucial factors in model performance. ImageNet ensures data quality through rigorous validation, with each image verified by an average of 3.3 annotators and achieving 95\% accuracy in its labels~\cite{deng2009imagenet}. This precise annotation enables models to learn accurate visual features and excel in image classification tasks. In the realm of large language models, GPT-3's training data underwent stringent cleaning and filtering, including deduplication, quality scoring based on document length and linguistic complexity, and content filtering for inappropriate content~\cite{brown2020language}. This high-quality data enables GPT-3 to generate coherent and accurate text. Furthermore, diversity is essential: ImageNet covers 1,000 object categories across various domains, while GPT-3's training data spans multiple languages, genres, and knowledge domains, providing rich linguistic knowledge and contextual understanding.

\paragraph{Data as the ceiling for model capabilities}
A model's capability depends on the knowledge it extracts from data, following empirically observed scaling laws. While increasing model parameters can enhance expressive power, without sufficient data, models cannot effectively utilize these parameters. DeepMind's research on the Chinchilla model demonstrated that under the same compute budget, a 70B parameter model trained on 1.4T tokens outperforms a 280B parameter model trained on 0.35T tokens, achieving a 30\% reduction in loss while using the same compute resources~\cite{hoffmann2022training}. This finding directly supports the notion that data acts as a ceiling for model capabilities. Additionally, Meta's research shows that while Llama 2 (70B) has 70 billion parameters, its performance largely benefits from training on 2T tokens of high-quality data, with particular emphasis on academic papers, code repositories, and books that enhance its reasoning capabilities~\cite{touvron2023llama}. These studies emphasize data's central role in model training and suggest that optimal model scaling requires a balanced increase in both parameters and training data.

\paragraph{Looking ahead}
From MNIST to ImageNet to GPT-3, advances in data scale, quality, and diversity have directly driven AI breakthroughs. Data remains the foundation of AI development, determining the upper limits of model capabilities. As we push the boundaries of LLM performance, the challenge of acquiring sufficient high-quality, diverse data becomes increasingly acute. Traditional data sources like the internet are showing signs of exhaustion, and concerns about data privacy and ownership are growing. This motivates the exploration of novel data acquisition approaches, such as leveraging edge devices and distributed data collection, which we will explore in subsequent sections. The future of LLMs may depend not just on scaling existing data sources, but on fundamentally rethinking how we collect, curate, and utilize data in AI training.

\subsection{Computing power: the engine of LLMs}

\paragraph{Early neural networks and CPU era}
Since the inception of neural networks, every breakthrough in the field of AI has been driven by the continuous improvement of computational power~\cite{thompson2020computational}. From the early multilayer perceptron (MLP) to the widely used large language models (LLM) today, the progress in computing power has always been a key engine for advancing AI.

As the prototype of neural networks, the MLP was initially used to solve linearly separable problems~\cite{rosenblatt1958perceptron}. Due to its relatively low computational demand, it could run on traditional CPU environments. However, as the complexity of neural network models increased and application scenarios expanded, computational requirements gradually rose. The emergence of Convolutional Neural Networks (CNN) and Recurrent Neural Networks (RNN) marked a surge in computational demands. CNN, through convolutional operations, effectively reduced the number of parameters, enhancing the computational efficiency of image processing tasks. Classic models such as LeNet~\cite{lecun1998gradient} and AlexNet~\cite{krizhevsky2012imagenet} achieved significant results in image classification, but this also led to a surge in computational resource demands. For example, AlexNet's victory in the 2012 ImageNet competition was made possible by using the NVIDIA GTX 580 GPU, which significantly boosted computational performance~\cite{krizhevsky2012imagenet}.

\paragraph{GPU and TPU revolution}
With the growing scale of neural network models, GPUs gradually became indispensable computing tools~\cite{raina2009large}. The parallel computing capabilities of GPUs greatly accelerated the training process of neural networks, particularly in the field of deep learning. Meanwhile, specialized hardware for deep learning, such as Tensor Processing Units (TPUs), emerged~\cite{jouppi2017datacenter}. Compared to GPUs, TPUs offer higher efficiency and lower power consumption when performing matrix operations and deep learning tasks~\cite{wang2019benchmarking}, making them the preferred hardware for training large-scale neural networks.

\paragraph{Transformer era and computational demands}
As computational resources continued to expand, the scale of neural network model training also grew. The introduction of the Transformer architecture~\cite{vaswani2017attention} revolutionized the field of natural language processing (NLP), especially with the launch of models like BERT~\cite{devlin2018bert} and the GPT series~\cite{brown2020language,openai2023gpt4}, which pushed NLP technology to new heights. However, the self-attention mechanism in the Transformer architecture has a computational complexity of $O(n^2)$, where n represents the sequence length~\cite{vaswani2017attention}. This means that as the model scale and sequence length increase, the required computational power grows exponentially. For example, training large language models like GPT-3~\cite{brown2020language} and GPT-4~\cite{openai2023gpt4} involves trillions of parameters and requires thousands of GPUs or TPU nodes to support the process. This immense computational demand not only places extremely high requirements on hardware, but also on computational frameworks, storage, and communication bandwidth, creating unprecedented challenges~\cite{patterson2021carbon}.

\paragraph{Computing power as the key driver}
Every leap in Artificial Intelligence has been driven by computational power~\cite{amodei2018ai}. From multilayer perceptrons to convolutional neural networks, and the introduction of the Transformer architecture, every innovation in models has been accompanied by an explosive growth in computational needs~\cite{thompson2020computational}. Particularly in the era of large language models, computational power is not only the foundational tool for model training but also the core driving force behind breakthroughs in AI performance~\cite{kaplan2020scaling}. The success of large-scale models like GPT-4 validates that AI progress almost entirely depends on the support of more powerful computational resources~\cite{hoffmann2022training}.

\section{Smartphone Data Volume Estimation}  
\label{app:smartphone_ethod}

In the absence of publicly available, granular data on per-user smartphone data generation patterns, we adopt a conservative estimation approach to approximate the total annual smartphone data volume. While this method necessarily involves simplifications, it provides a robust lower-bound approximation that is sufficient to support our core arguments without compromising the validity of our conclusions.

\textbf{Data volume estimation per smartphone}: Based on industry reports  \cite{counterpoint_smartphone_2021}, the average smartphone storage capacity reached 100 GB in 2020. To ensure a conservative estimate, we assume that only 1\% of this storage capacity (equivalent to approximately 1 GB per smartphone) is actively used for data generation and storage, including local images, video information, and other types of user-generated content. This assumption aligns with baseline usage scenarios while intentionally underestimating actual data utilization.

\textbf{Number of smartphones}: The growth of the number of smartphone users is an important basis for estimating the total amount of data. For this, we have referred to data from market research institutions \cite{bankmycell_smartphone_2023}, which includes trends in changes to the number of smartphone users over time.

Based on the above statistical data, the total annual smartphone data volume \( D_{\text{total}} \) is calculated using the following formula:  
\begin{equation}  
    D_{\text{total}}  (\text{EB}) = N_{\text{users}} \times 1 \, \text{GB/user} \times 10^{-3}  \, (\text{conversion from GB to EB}),
\end{equation}  
where \( N_{\text{users}} \) represents the global smartphone user base in billions.  

Substituting \( N_{\text{users}} = 8.0 \times 10^9 \) (representing 8 billion users) into Equation (1):  
\begin{equation*}  
    D_{\text{total}} = 8.0 \, \text{GB/user} \times 10^{-3} = 8.0 \, \text{EB}.
\end{equation*}  

Our purpose is to establish a defensible lower bound for analysis. Even under these stringent assumptions, the derived volumes remain orders of magnitude higher than synthetic or centralized datasets, thereby reinforcing the strategic importance and value of edge-generated data. This conservative estimation underscores the critical need for scalable solutions capable of managing and leveraging such vast quantities of distributed data effectively.

\begin{table}[!ht]
\centering
\caption{Trends in Smartphone Shipments and Compute Power. (Data source: \cite{canalys2025}).}
\label{tab:chip_total}
\resizebox{\linewidth}{!}{%
\begin{tabular}{cccc}
\hline
\textbf{Company} & \textbf{Shipments (Million units)} & \textbf{Chip Performance Range (TFLOPS)} & \textbf{Total Compute Power Contribution (EFLOPS)} \\
\hline
\multicolumn{4}{c}{\textbf{2020}} \\ \hline
Samsung (20\%) & 255.5 & 1.20--1.53 & 349 \\
Apple (16\%)   & 207.2 & 0.65 & 135 \\
Xiaomi (12\%)  & 149.6 & 0.24--1.20 & 108 \\
OPPO (9\%)    & 119.4 & 0.24--1.20 & 86 \\
vivo (9\%)    & 112.6 & 0.24--1.20 & 81 \\
Others (33\%)  & 420.5 & 0.04--0.24 & 59 \\ \hline
\multicolumn{4}{c}{\textbf{Overall: Shipments = 1265 Million, Compute Power = 817 EFLOPS}} \\ \hline \hline

\multicolumn{4}{c}{\textbf{2021}} \\ \hline
Samsung (20\%) & 274.5 & 1.42--1.72 & 430 \\
Apple (17\%)   & 230.1 & 1.71--1.94 & 420 \\
Xiaomi (14\%)  & 191.2 & 0.82--1.74 & 240 \\
OPPO (11\%)    & 145.1 & 0.82--1.74 & 180 \\
vivo (10\%)    & 129.9 & 0.82--1.74 & 160 \\
Others (28\%)  & 379.4 & 0.27--0.82 & 207 \\ \hline
\multicolumn{4}{c}{\textbf{Overall: Shipments = 1350 Million, Compute Power = 1637 EFLOPS}} \\ \hline \hline

\multicolumn{4}{c}{\textbf{2022}} \\ \hline
Samsung (22\%) & 257.9 & 0.49--2.01 & 322 \\
Apple (19\%)   & 232.2 & 1.79 & 416 \\
Xiaomi (13\%)  & 152.7 & 1.01--3.49 & 351 \\
OPPO (10\%)    & 113.4 & 1.01--3.49 & 261 \\
Transsion (6\%)     & 73.1 & 0.24--0.98 & 44.6 \\
Others (31\%)  & 364.1 & 0.84--1.31 & 393 \\ \hline
\multicolumn{4}{c} {\textbf{Overall: Shipments = 1193 Million, Compute Power = 1788 EFLOPS}} \\ \hline \hline

\multicolumn{4}{c}{\textbf{2023}} \\ \hline
Apple (20\%)   & 229.1 & 2.15 & 493 \\
Samsung (20\%) & 225.5 & 2.01--2.77 & 539 \\
Xiaomi (13\%)  & 146.1 & 2.15--3.99 & 449 \\
OPPO (9\%)     & 100.7 & 2.15--3.99 & 309 \\
Transsion (8\%)  & 92.6 & 0.24--1.31 & 72 \\
Others (30\%)  & 347.9 & 0.24--2.15 & 416 \\
\hline
\multicolumn{4}{c}{\textbf{Overall: Shipments = 1142 Million, Compute Power = 2278 EFLOPS}} \\ \hline \hline

\multicolumn{4}{c}{\textbf{2024}} \\ \hline
Apple (18\%)   & 225.9 & 1.91--2.29 & 474 \\
Samsung (18\%) & 222.9 & 3.38--3.41 & 758 \\
Xiaomi (14\%)  & 168.6 & 3.38--4.95 & 703 \\
Transsion (9\%)  & 106.7 & 0.05--0.67 & 38  \\
OPPO (8\%)   & 103.6 & 3.38--4.95 & 432 \\
Others (33\%)  & 395.4 & 0.05--1.72 & 352 \\
\hline
\multicolumn{4}{c}{\textbf{Overall: Shipments = 1223 Million, Compute Power = 2758 EFLOPS}} \\ \hline
\end{tabular}%
}
\end{table}

\section{Estimation of Smartphone Total Computational Power}  
\label{app:total_computation}

To assess the (ideally) aggregate computational capabilities of smartphones globally, we estimate the total computing power, given the current lack of comprehensive statistical data in this domain. Our approach leverages two key data sources: the annual worldwide shipment volumes for major smartphone brands, and the computational performance specifications of mobile processors deployed in their devices during each corresponding year. The complete data underlying our analysis is presented in Table~\ref{tab:chip_total}, which provides a detailed breakdown by manufacturer and time period.
For quantitative analysis, we formulated a mathematical model to calculate the total computing power. Specifically, for any given year, we compute the aggregate computational capacity ($C_{\text{total}}$) by summing the contributions from each smartphone manufacturer ($i$). Each manufacturer's contribution is determined by multiplying their total device shipments ($N_i$) by the average computing power of their mobile processors ($P_i$) for that year, expressed formally as:

\begin{equation}
    C_{\text{total}} = \sum_{i} N_i \cdot P_i
\end{equation}

This formulation enables us to systematically track the evolution of distributed computing power across the smartphone ecosystem while accounting for both market share dynamics and technological advancement in mobile processors. By maintaining conservative estimates for processor capabilities and focusing on verified shipment data, our analysis provides a reliable lower bound for the total computational resources available through smartphones.

\section{Small Language Model (SLM) Architectures and Training Methods}
\label{app:slm_architectures_training}

Table~\ref{tab:slm_architectures_training} presents a comprehensive overview of the Small Language Model (SLM) landscape, categorized by architectures and training methodologies, according to \cite{wang2024comprehensive}. The table is organized into two main categories: (I) Transformer-Based Models, which represent the dominant architecture in current SLMs, and (II) Alternative Architecture Models, which explore novel approaches to achieve efficiency. The Transformer-Based section is further divided into models pre-trained from scratch, models derived from larger LLMs through knowledge distillation, and models created through various compression techniques (pruning, quantization, etc.). The Alternative Architecture section showcases emerging approaches like State Space Models (Mamba, Hymba), recurrent architectures (RWKV, xLSTM), and traditional encoder-decoder or encoder-only designs.

\begin{table}[!ht]
    \caption{Small Language Model (SLM) Architectures and Training Methods}
    \label{tab:slm_architectures_training}
    \tiny
    \resizebox{\textwidth}{!}{
    \begin{tabularx}{\textwidth}{p{2.5cm}p{1cm}p{1cm}ccp{2.2cm}p{2cm}}
    \toprule
    \textbf{Model} & \textbf{Sizes} & \textbf{Architecture} & \textbf{From Scratch} & \textbf{From LLMs} & \textbf{Training Method} & \textbf{Datasets} \\
    \midrule
    
    \multicolumn{7}{l}{\textbf{\textit{I. Transformer-Based Models}}} \\
    \midrule
    
    \multicolumn{7}{l}{\textit{I.A. Pre-Trained from Scratch}} \\
    \addlinespace[0.5ex]
    PhoneLM \cite{yi2024phonelm} & 0.5B; 1.5B & Transformer & \checkmark & & Pre-training & DCLM-baseline \cite{li2024datacomp}, StarCoderData \cite{li2023starcodersourceyou} \\
    Llama 3.2 \cite{llama3.2} & 1B; 3B & Transformer & \checkmark & & Pre-training, SFT, RLHF, DPO & Not released (9T tokens) \\
    Qwen 1/1.5/2/2.5 \cite{yang2024qwen2,bai2023qwentechnicalreport} & 0.5B-7B & Transformer & \checkmark & & Pre-training & Not released \\
    Gemma \cite{team2024gemma} & 2B; 7B & Transformer & \checkmark & & Pre-training & Unknown \\
    SmolLM2 \cite{allal2025smollm2smolgoesbig} & 135M-1.7B & Transformer & \checkmark & & Pre-training & SmolLM corpus \cite{benallal2024smollmcorpus} \\
    H2O-Danube3 \cite{pfeiffer2024h2o} & 500M; 4B & Transformer & \checkmark & & Pre-training (multi-stage) & Unknown \\
    MiniCPM \cite{hu2024minicpm} & 1.2B; 2.4B & Transformer & \checkmark & & Pre-training & Dolma \cite{dolma}, C4 \cite{raffel2020exploring} \\
    CT-LLM \cite{du2024chinese} & 2B & Transformer & \checkmark & & Pre-training & MAP-CC \\
    OLMo \cite{groeneveld2024olmo} & 1B; 7B & Transformer & \checkmark & & Pre-training & Dolma \cite{dolma} (multiple sources) \\
    TinyLlama \cite{zhang2024tinyllamaopensourcesmalllanguage} & 1B & Transformer & \checkmark & & Pre-training & SlimPajama \cite{cerebras2023slimpajama} \\
    Phi-series \cite{abdin2024phi,javaheripi2023phi} & 1.3B-6.6B & Transformer & \checkmark & & Pre-training & CodeTextBook \cite{gunasekar2023textbooksneed} \\
    OpenELM \cite{mehta2024openelm} & 270M-3B & Transformer & \checkmark & & Pre-training & RefinedWeb \cite{penedo2023refinedweb}, PILE \cite{gao2020pile} \\
    MobiLlama \cite{thawakar2024mobillama} & 0.5B; 0.8B & Transformer & \checkmark & & Pre-training & LLM360 Amber \\
    MobileLLM \cite{liu2024mobilellm} & 125M; 350M & Transformer & \checkmark & & Pre-training & Unknown (1T tokens) \\
    \addlinespace[0.5ex]
    
    \midrule
    \multicolumn{7}{l}{\textit{I.B. Derived from Larger Models}} \\
    \addlinespace[0.5ex]
    MINITRON \cite{muralidharan2024compact} & 4B & Transformer & & \checkmark & Distillation, Pruning & 8T tokens from Nemotron-4 \\
    Orca/Orca 2 \cite{mitra2023orca,mukherjee2023orca} & 7B; 13B & Transformer & & \checkmark & Distillation & Orca 2 dataset, FLAN-v2 \cite{longpre2023flan} \\
    MINIMA \cite{zhang2023towards} & 3B & Transformer & & \checkmark & Distillation (from Llama-2-7B) & Pile \cite{gao2020pile}, Wudao \\
    Dolly-v2 \cite{DatabricksBlog2023DollyV2} & 3B; 7B & Transformer & & \checkmark & Instruction tuning (from Pythia) & Databricks-dolly-15k \\
    LaMini-LM \cite{wu2023lamini} & 61M-7B & Transformer & & \checkmark & Distillation & LaMini instruction dataset \\
    \addlinespace[0.5ex]

    \midrule
    \multicolumn{7}{l}{\textit{I.C. Model Compression Approaches}} \\
    \addlinespace[0.5ex]
    SparseGPT \cite{frantar2023sparsegpt} & Various & Transformer & & \checkmark & Unstructured Pruning & Not applicable \\
    Wanda \cite{sun2023wanda} & Various & Transformer & & \checkmark & Unstructured Pruning & Not applicable \\
    LoRAPrune \cite{zhang2023loraprune} & Various & Transformer & & \checkmark & Unstructured Pruning & Not applicable \\
    ShortGPT \cite{men2024shortgpt} & Various & Transformer & & \checkmark & Structured Pruning & Not applicable \\
    BitNet/BitNet b1.58 \cite{wang2023bitnet,ma2024era} & Various & Transformer & & \checkmark & Quantization (QAT) & Not applicable \\
    QLoRA \cite{dettmers2023qlora} & Various & Transformer & & \checkmark & Quantization, Low-Rank & Various fine-tuning datasets \\
    SqueezeLLM \cite{kim2024squeezellm} & Various & Transformer & & \checkmark & Quantization (PTQ) & Not applicable \\
    \midrule
    
    \multicolumn{7}{l}{\textbf{\textit{II. Alternative Architecture Models}}} \\
    \midrule
    
    \addlinespace[0.5ex]
    Mamba \cite{gu2023mamba} & 125M-1.3B & Mamba & \checkmark & & Pre-training & Pile \cite{gao2020pile} \\
    Rene \cite{Rene} & 1.3B & Mamba & \checkmark & & Pre-training & Dolma-1.7 \cite{dolma} \\
    Zamba2 \cite{glorioso2024zamba2} & 2.7B & Mamba & \checkmark & & Pre-training & Not specified \\
    Hymba \cite{dong2024hymba} & 125M-1.5B & Hymba & \checkmark & & Pre-training & DCLM-Baseline \cite{li2024datacomp} \\
    xLSTM \cite{beck2024xlstm} & 125M-1.3B & xLSTM & \checkmark & & Pre-training & SlimPajama \cite{cerebras2023slimpajama} \\
    RWKV \cite{peng-etal-2023-rwkv} & 169M-14B & RNN & \checkmark & & Pre-training & Pile \cite{gao2020pile} \\
    \addlinespace[0.5ex]
    
    \addlinespace[0.5ex]
    Specialized FlanT5 \cite{fu2023specializing} & 250M-3B & Encoder-Decoder & & \checkmark & Instruction Tuning & GSM8K \cite{cobbe2021gsm8k} \\
    FlanT5 \cite{chung2022scaling} & 80M-3B & Encoder-Decoder & & \checkmark & Instruction Tuning & Muffin, T0-SF, SNI and CoT \\
    T5 \cite{raffel2020exploring} & 60M-3B & Encoder-Decoder & \checkmark & & Pre-training & C4 \cite{raffel2020exploring} \\
    \addlinespace[0.5ex]
    
    \addlinespace[0.5ex]
    DistilBERT \cite{sanh2019distilbert} & 66M & Encoder-only & & \checkmark & Distillation (from BERT) & Wikipedia, BookCorpus \\
    TinyBERT \cite{jiao2020tinybert} & 14.5M & Encoder-only & & \checkmark & Distillation (from BERT) & Wikipedia, BookCorpus \\
    ALBERT \cite{lan2020albert} & 12M-18M & Encoder-only & \checkmark & & Pre-training (parameter sharing) & Wikipedia, BookCorpus \\
    \bottomrule
    \end{tabularx}
    }
    \end{table}

    This classification showcases the architectural innovations and training methodologies that are driving the SLM field forward, providing essential technical foundations for deploying powerful AI capabilities on resource-constrained edge devices. By documenting various model sizes, training corpora, and development techniques, the table offers a comprehensive overview of cutting-edge approaches that enable sophisticated language processing directly on end-user devices. These advancements represent critical building blocks for the next generation of on-device AI systems that can operate efficiently without constant cloud connectivity while still delivering robust performance across diverse applications.

\newpage
\section{Distributed Collaborative Frameworks}
\label{app:distributed_collaborative_frameworks}
Distributed collaborative frameworks enable the deployment, training, and fine-tuning of language models across multiple devices or servers. Table \ref{tab:framework_comparison} presents a comparison of prominent frameworks in this domain. These frameworks can be broadly categorized into three types: cloud-based platforms that offer centralized resources for distributed computing, federated learning systems that enable training across decentralized data sources while preserving privacy, and fully decentralized frameworks that distribute computation across peer nodes. Some frameworks like Neurosurgeon \cite{kang2017neurosurgeon}, MoE$^2$ \cite{jin2025moe2}, Edgent \cite{li2018edgent}, and Galaxy \cite{ye2024galaxy} focus on collaborative inference by partitioning models between edge devices and servers. Others, such as FedDAT \cite{chen2024feddat}, FedFM \cite{yang2025federated}, FedPETuning \cite{zhang2023fedpetuning}, and Photon \cite{sani2024photon}, specialize in federated fine-tuning of large language models while maintaining data privacy. These frameworks are essential for enabling efficient deployment of language models in resource-constrained environments and for scenarios requiring privacy preservation or operation in disconnected settings.

\begin{table}[h!]
    \centering
    \resizebox{\linewidth}{!}{%
    \begin{tabular}{l*{7}{c}}
    \toprule
    & \multicolumn{4}{c}{\textbf{Distributed Capabilities}} & & & \\
    \cmidrule(lr){2-5}
    \textbf{Framework} & \textbf{Inference} & \textbf{Training} & \textbf{Pretraining} & \textbf{Fine-tuning} & \textbf{Type} & \textbf{Privacy} & \textbf{License} \\
    \midrule
    exo-explore/exo \cite{exo} & \checkmark &  &  &  & Decentralized &  & MIT \\
    Together AI \cite{together} & \checkmark & \checkmark & \checkmark & \checkmark & Cloud &  & Commercial \\
    FLock Platform \cite{flock} &  & \checkmark &  & \checkmark & Federated, Blockchain & \checkmark & Apache 2.0  \\
    OpenDiloco \cite{OpenDiLoCo} &  & \checkmark & \checkmark &  & Decentralized &  & Apache 2.0 \\
    FederatedScope \cite{federatedscope} &  & \checkmark &  & \checkmark & Federated & \checkmark & Apache 2.0 \\
    FedML \cite{fedml} &  & \checkmark &  & \checkmark & Federated & \checkmark & Apache 2.0 \\
    Flower \cite{Flower} &  & \checkmark &  & \checkmark & Federated & \checkmark & Apache 2.0 \\
    FATE-LLM \cite{fan2023fate} &  & \checkmark &  & \checkmark & Federated & \checkmark & Apache 2.0 \\
    FedLLM \cite{ye2024fedllm} &  & \checkmark & \checkmark & \checkmark & Federated & \checkmark & CC BY-NC 4.0 \\
    \bottomrule
    \end{tabular}%
    }
    \caption{Comparison of Distributed Machine Learning Frameworks}
    \label{tab:framework_comparison}
\end{table}

\end{document}